\documentclass[11pt,preprint]{aastex}

\begin{document}

\title{On the Importance of Conceptual Thinking Outside the Simulation
Box}

\author{Abraham Loeb\\Institute for Theory \& Computation\\
Harvard University\\60 Garden St., Cambridge, MA 02138}

\begin{abstract}

Any ambitious construction project requires architects for its design
and engineers who apply the design to the real world. As scientific
research shifts towards large groups which focus on the engineering
aspects of linking data to existing models, architectural skills are
becoming rare among young theorists. Senior researchers should mentor
qualified students and postdocs to think creatively about the big
picture without unwarranted loyalty to ancient blueprints from past
generations of architects.

\end{abstract}

{\it ~\\``If you always do what you always did, you will always get what you
always got.''}

\noindent
{\it Albert Einstein}

\bigskip
\bigskip
\bigskip

Too few theoretical astrophysicists are engaged in tasks that go
beyond the refinement of details in a commonly accepted paradigm. It
is far more straightforward today to work on these details than to
review whether the paradigm itself is valid.  While there is much work
to be done in the analysis and interpretation of experimental data,
the unfortunate by-product of the current state of affairs is that
popular, mainstream paradigms within which data is interpreted are
rarely challenged.  Most cosmologists, for example, lay one brick of
phenomenology at a time in support of the standard
(inflation$+\Lambda+$Cold-Dark-Matter) cosmological model, resembling
engineers that follow the blueprint of a global construction project,
without pausing to question whether the architecture of the project
makes sense when discrepancies between expectations and data are
revealed.

The problem with researchers focusing on the engineering aspects of a
prevailing paradigm rather than on questioning its foundation is that
their efforts to advance scientific knowledge are restricted to a
conservative framework. For example, cosmological data is often
analyzed in the conservative mindset of a community-wide effort to
reduce the error budget on measurements of the standard cosmological
parameters. The cosmology community has become so conservative that
when a discrepancy was identified recently between different methods
for deriving $\Omega_m$ and $\sigma_8$ (using microwave background
anisotropies sourced at redshift $z\sim 10^3$ and cluster abundance
measurements at $z\lesssim 1$) in the latest data from the Planck
satellite, the mundane possibility of assigning a mass of 0.2eV to
neutrinos is being debated as a wild deviation from the mainstream
comfort zone.  Deviations of this magnitude were trivial for previous
generations of cosmologists who were debating the underlying physical
principles that control the Universe while entertaining dramatic
excursions from conservative guidelines.

Conservatism is possible today because we now have a standard
cosmological model, which was not in place several decades ago. It is
nevertheless disappointing to see conformism among young cosmologists
who were born into the standard paradigm and are supposed to be least
biased by prejudice. The experience resembles seeing the children of
hippies from the 1960s transform surprisingly quickly into
establishment executives. This phenomenon has its obvious reasons. The
unfortunate reality of young astrophysicists having to spend their
most productive years in lengthy postdoctoral positions without job
security promotes conformism, as postdocs aim to improve their chance
of getting a faculty job by supporting the prevailing paradigm of
senior colleagues who serve on selection committees. Ironically, one
might argue that an opposite strategy of choosing innovative projects
should improve the job prospects of a candidate, since it would
separate that candidate from the competing crowd of indistinguishable
applicants that selection committees are struggling through.

The orthodoxy exhibited by young cosmologists today raises eyebrows
among some of the innovative ``architects'' who participated in the
design of the standard model of cosmology.  Too soon after the
enigmatic ingredients of the standard cosmology were confirmed
observationally, they acquired the undeserving status of an absolute
truth in the eyes of beginning cosmologists.

One would have naively expected scientific activity to be open minded
to critical questioning of its architectural design, but the reality
is that conservatism prevails within the modern academic
setting. Orthodoxy with respect to mainstream scientific dogmas does
not lead to extreme atrocities such as burning at the stake for heresy
but it propagates other collective punishments, such as an unfair
presentation of an innovative idea at conferences, bullying, and
drying up of resources for innovative thinkers.

The problem is exacerbated by the existence of large groups with a
rigid, prescribed agenda and a limited space for innovation when
unexpected results emerge.  In large groups, such as the Planck or
SDSS-III collaborations, young cosmologists often decide not to
challenge the established paradigm because other group members, and
particularly senior scientists who are considered experts on the
issue, accept this paradigm.  If hundreds of names appear on the
author list of a paper, the vast majority of them have limited space
for maneuvers -- like a dense swarm of fish in a small aquarium.  The
fact that letters of recommendation are written by few group leaders
adds pressure to conform to mainstream ideas.  True, some scientific
goals require a large investment of funds and research time, but under
these circumstances efforts should be doubled to maintain innovative
challenges to mainstream thinking. It is imperative that {\it
individual} scientists feel comfortable expressing critical views in
order for the truth to ultimately prevail.

And there is no better framework for critically challenging a
prevailing dogma than at the architectural ``blueprint''
level. Sometimes, a crack opens in a very particular corner of a dogma
due to a localized discrepancy with data. But conceptual anomalies are
much more powerful since they apply to a wide range of phenomena and
are not restricted to a corner of parameter space.  Albert Einstein's
thought experiments are celebrated stepping stones that identified
earlier conceptual anomalies and paved the path to the theories of
Special and General Relativity in place of the fixed spacetime
blueprint of Isaac Newton. And before Galileo Galilei came up with his
conceptual breakthrough, it was standard to assume that heavy objects
fall faster than light objects under the influence of gravity.

Conceptual work is often undervalued in the minds of those who work on
the details. A common misconception is that the truth will inevitably
be revealed by working out the details. But this misses the biggest
blunder in the history of science: that the accumulation of details
could be accommodated within any prevailing paradigm by tweaking and
complicating the paradigm. A classic example is Ptolemaic cosmology, a
theory of epicycles for the motion of the Sun and planets around the
Earth that survived empirical scrutiny for longer than it deserved. A
modern analog is the conviction shared by mainstream cosmologists that
the matter density in the Universe equals the critical value,
$\Omega_m=1$. This notion dominated in the 1980s and the early 1990s
for nearly two decades after inflation was proposed, a period during
which discrepancies between data and expectations were assigned to an
unknown ``bias parameter'' of galaxies even when data on the mass to
light ratio on large spatial scales indicated clearly that
$\Omega_m\sim 0.3$ as we now know to be the case. Today, when
discrepancies between the observed distribution of dark matter inside
galaxies and theoretical expectations for cold dark matter halos are
revealed, these discrepancies are commonly brushed aside as being due
to ``uncertain baryonic physics'' even in dwarf galaxies where the
baryons make a negligible contribution to the overall mass budget.
Similarly, when a hemispherical asymmetry in the power spectrum of
temperature fluctuations of the cosmic microwave background was
reported a decade ago, it was quickly dismissed by mainstream
cosmologists; this anomaly is now confirmed by the latest data from
the Planck satellite but still viewed as an unlikely ($<1$\%
probability) realization of the sky in the standard,
statistically-isotropic cosmology. Given that progress in physics was
historically often motivated by experimental data, it is surprising to
see that a speculative concept with no empirical basis, such as the
``multiverse'', gains traction among some mainstream cosmologists,
whereas a data-driven hypothesis, such as dark matter with strong
self-interaction (to remedy galactic scale discrepancies of
collisionless dark matter), is much less popular.

The most efficient way to simplify the interpretation of data is to
work at the meta-level of architecture, similarly to the helio-centric
interpretation developed by Copernicus in the days when the Ptolemaic
theory ruled. An engineering project which aims to study the strength
of rods and bricks might never lead to a particularly elegant
blueprint for the building in which these ingredients are finally
embedded. In the scientific quest for the truth, we also need
architects (theory makers), not just engineers (theory
implementers/testers).

Some argue that architects were only needed in the early days of a
field like cosmology when the fundamental building blocks of the
standard model, e.g., the inflaton, dark matter and dark energy, were
being discovered. As fields mature to a state where quantitative
predictions can be refined by detailed numerical simulations, the
architectural skills are no longer required for selecting a winning
world model based on comparison to precise data. Ironically, the
example of cosmology demonstrates just the opposite. On the one hand,
we measured various constituents of our Universe to two significant
digits and simulated them with accurate numerical codes. But at the
same time, we do not have a fundamental understanding of the nature of
the dark matter or dark energy nor of the inflaton.  In searching for
this missing knowledge, we need architects who could suggest to us
what these constituents might be in light of existing data and which
observational clues should be searched for.  Without such clues, we
will never be sure that inflation really took place or that dark
matter and dark energy are real and not ghosts of our imagination due
to a modified form of gravity.

It is true that numerical codes allow us to better quantify the
consequences of a prevailing paradigm, but systematic offsets between
these results and observational data cannot be cured by improving the
numerical precision of these codes. The solutions can only be found
outside the simulation box through conceptual thinking.  For example,
observations of X-ray clusters are currently being calibrated at the
percent level using hydrodynamic simulations as a precise tool for
inferring cosmological parameters.  However, important physical
effects that may influence the temperature and density distribution of
the hot intracluster gas, such as heat conduction, thermal
evaporation, and electron-ion temperature equilibration, are left out
of these simulations despite the fact that the Coulomb mean-free-path
of protons in the outskirts of clusters is comparable to the cluster
radius.

Conceptual thinkers are becoming an extinct species in the landscape
of present-day astrophysics. I find this trend worrisome for the
health of our scientific endeavor. In my view, it is our duty to
encourage young researchers to think critically about prevailing
paradigms and to come up with simplifying conceptual remedies that
take us away from our psychological comfort zone but closer to the
truth.

\bigskip
\bigskip
\bigskip

\acknowledgements I thank Richard Ellis, Ido Liviatan, Robb Scholten
and Josh Winn for helpful comments on the manuscript.

\end{document}